\documentclass[12pt]{article}

\usepackage{scicite}

\usepackage{times}

\usepackage{amsmath}
\usepackage{amsfonts}
\usepackage{amssymb}
\usepackage{graphicx}

\usepackage{xcolor}
\definecolor{darkblue}{rgb}{0,0,0.6}
\definecolor{darkred}{rgb}{0.6,0,0}
\newcommand{\Jij}{J_{ij}}

\newcommand{\Hoff}{H^-}
\newcommand{\Hon}{H^+}

\newenvironment{sciabstract}{%
\begin{quote} \bf}
{\end{quote}}

\title{Multiperiodic orbits from interacting soft spots in cyclically-sheared amorphous solids}

\author
{Nathan C.\ Keim,${}^{1,2\ \ast\ddag}$ Joseph D.\ Paulsen,${}^{3,4\ \dag\ddag}$\\
\\
\normalsize{${}^{1}$Department of Physics, Pennsylvania State University, University Park, PA 16802, USA}\\
\normalsize{${}^{2}$Department of Physics, California Polytechnic State University, }\\
\normalsize{San Luis Obispo, CA 93407, USA}\\
\normalsize{${}^{3}$Department of Physics, Syracuse University, Syracuse, NY 13244, USA}\\
\normalsize{${}^{4}$BioInspired Syracuse: Institute for Material and Living Systems, }\\
\normalsize{Syracuse University, Syracuse, NY 13244, USA}\\
\normalsize{$^\ast$To whom correspondence should be addressed; E-mail:  keim@psu.edu.} \\
\normalsize{$^\dag$To whom correspondence should be addressed; E-mail: jdpaulse@syr.edu.} \\
\normalsize{$^\ddag$Equal contributions to this work.}
}

\date{}

\begin{document} 


%\baselineskip24pt


\maketitle


% The abstract should be a single paragraph, not to exceed 250 words and ideally closer to 200, written in plain language that a general reader can understand. It should include
% An opening sentence that states the question/problem addressed by the research AND
% Enough background content to give context to the study AND
% A brief statement of primary results AND
% A short concluding sentence.
% Do not include citations or undefined abbreviations in the abstract. Any abbreviations that appear in the title should be defined in the abstract.

\begin{sciabstract}
  When an amorphous solid is deformed cyclically, it may reach a steady state in which the paths of constituent particles trace out closed loops that repeat in each driving cycle. A remarkable variant has been noticed in simulations where the period of particle motions is a multiple of the period of driving, but the reasons for this behavior have remained unclear. 
Motivated by the mesoscopic features of displacement fields in experiments on jammed solids, we propose and analyze a simple model of interacting soft spots---locations where particles rearrange under stress, and that resemble two-level systems with hysteresis. 
We show that multiperiodic behavior can arise among just three or more soft spots that interact with each other, but in all cases it requires frustrated interactions, illuminating this otherwise elusive type of interaction. 
We suggest directions for seeking this signature of frustration in experiments, and we describe how to achieve it in designed systems. 
\end{sciabstract}

% In setting up this template for *Science Advances* papers, both
% the \section* command and the \paragraph* command are used for topical
% divisions.  Which you use will of course depend on the type of paper
% you're writing.  Review Articles tend to have displayed headings, for
% which \section* is more appropriate; Research Articles, when they have
% formal topical divisions at all, tend to signal them with bold text
% that runs into the paragraph, for which \paragraph* is the right
% choice.  Either way, use the asterisk (*) modifier, as shown, to

\section*{Introduction}
\begin{figure}[t]
\centering
\includegraphics[width=0.6\textwidth]{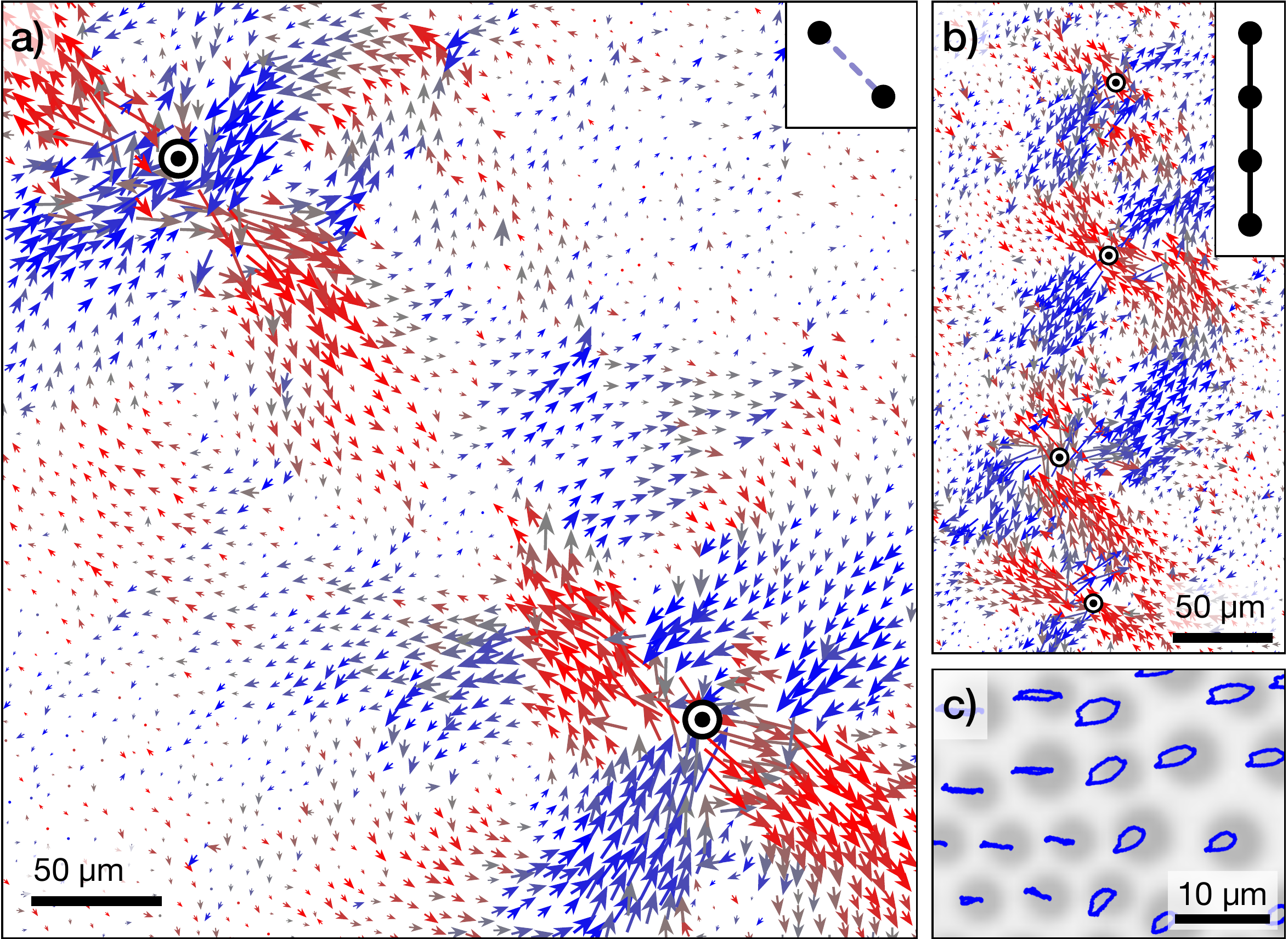}
%stz-interactions.pdf}
\caption{
\textbf{Interacting soft spots and periodic orbits in experiments on a cyclically-sheared 2D jammed solid.} 
\textbf{(a)} Particle displacements around two rearranging soft spots (approximate centers marked with $\odot$), undergoing horizontal shear. 
Colors denote displacements along the two principal axes of shear. 
The displacements oppose each other at the center of the panel, suggesting a frustrated interaction. \textit{Inset:} Schematic of the frustrated interaction (dashed line).
\textbf{(b)} Displacements around a group of several soft spots, suggesting cooperative interactions. 
\textit{Inset:} Schematic of cooperative interactions (solid lines).
\textbf{(c)} Steady-state particle paths, which are closed with the same period as the driving. Multiperiodic paths would have a longer period. Background: experimental micrograph.
}
\label{fig:1}
\end{figure}

{A} solid with perfectly elastic behavior deforms reversibly, in the sense that all material points return to their initial positions when a load is removed. 
Remarkably, some amorphous solids may be prepared in a reversible \textit{plastic} state, wherein loading the material in one direction changes its structure through many microscopic events, but loading it in the reverse direction precisely undoes these changes~\cite{Regev13,Keim13,Nagamanasa14,Keim14,Regev15}. 
Each microscopic event is localized to a soft spot \cite{Manning11} or shear-transformation zone (STZ) \cite{Falk11} (Fig.~\ref{fig:1}a), which resembles a two-level system that switches under forward and reverse shear \cite{Falk11,Keim14,Mungan19,Keim20}. 

Recent simulations using athermal quasistatic shear have revealed an even more remarkable behavior in which the period of particle motions is a \emph{multiple} of the period of driving \cite{Schreck13,Regev13}, reminiscent of the familiar action of a retractable pen. Such ``multiperiodic'' behavior may sound quite tenuous, given the daunting number of mechanically-stable configurations and transitions in a packing of even a modest size. Nevertheless, multiperiodicity has been observed in molecular dynamics simulations of amorphous solids in two and three dimensions, for several kinds of particle interactions~\cite{Schreck13,Regev13,Royer15,Kawasaki16,Lavrentovich17,Nagasawa19,Yeh20}. However, the mechanism for this behavior has remained unclear, even as it seems to be associated with an unjamming transition as the confining pressure is decreased~\cite{Lavrentovich17}. %---indeed, it has been unclear whether it can be explained in terms of soft spots, or is strictly an emergent behavior of many interacting particles.

\begin{figure}[t]
\centering
\includegraphics[width=0.6\textwidth]{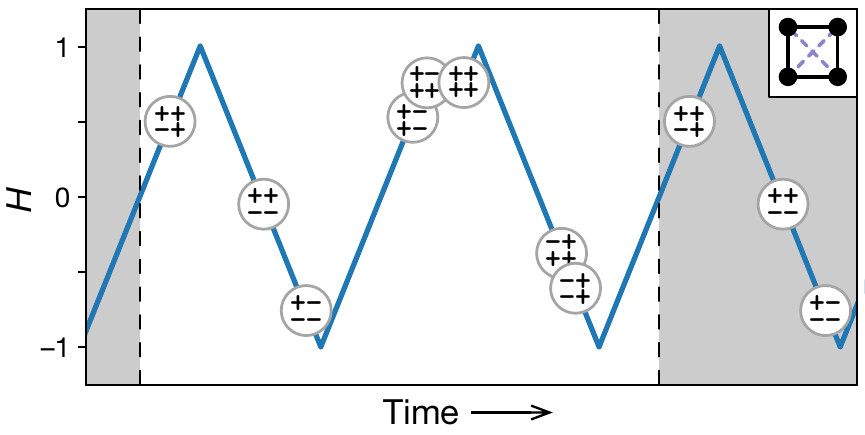}
\caption{
\textbf{Example of a $T=2$ orbit in our model of interacting hysterons. 
} 
\textit{Inset:} Arrangement of the four hysterons with a mixture of cooperative and frustrated interactions (solid and dashed lines, respectively). 
\textit{Main:} Each large circle represents a state of the system, and is placed at the value of external field $H$ at which the system reaches that state. The close pair of states near $H=1$ constitute an avalanche. The time axis indicates only the sequence of events, since the simulation is quasistatic. 
}
\label{fig:2}
\end{figure}

Here we show how multiperiodicity can arise in a simplified coarse-grained model of interacting soft spots (Fig.~\ref{fig:2}). 
We identify how the prevalence of multiperiodicity depends on the spatial arrangement of the soft spots, and we show how to design the behavior on demand. 
In all cases, the multiperiodic orbits are made possible by frustrated interactions in our model. 
Our results show that frustrated interactions between soft spots must be considered
as an important counterpart to the cooperative interactions that are used to explain avalanches near the yielding transition \cite{Maloney04, Regev15, Nicolas18}.

\section*{Results}
While experiments have not yet observed multiperiodic behavior, they exhibit the microscopic phenomenology we wish to distill into our model. Figure~\ref{fig:1}a shows a displacement field from an experiment with two nearby soft spots (see Materials and Methods for details). Each has the characteristics of an Eshelby inclusion---a small region of plastic deformation that is coupled to a quadrupolar elastic deformation of the surrounding material. This extended deformation induces or inhibits the rearrangement of other nearby soft spots, depending on their relative placement~\cite{Maloney04,Chikkadi11,Mungan19}. For example, Fig.~\ref{fig:1}a is suggestive of a frustrated interaction, whereas the arrangement in Fig.~\ref{fig:1}b suggests cooperative interactions. 
%These observations form the basis of our coarse-grained model. 

Our jumping-off point is to consider the possible behaviors of compact collections of $N$ soft spots by modeling them as interacting hysteretic elements, or ``hysterons'' \cite{Mungan19, Keim20}. A hysteron has two possible states, $s_i = \pm 1$; it transitions to the ``$+$'' state when the local field---equal to the instantaneous global strain field $H$ plus neighbor interactions---reaches a fixed threshold $\Hon_i$. Likewise, it transitions to the ``$-$'' state at a fixed threshold $\Hoff_i < \Hon_i$. To model the disorder of such packings, these thresholds are set as $\Hon_i = h_i + u_i$, $\Hoff_i = h_i - u_i$, where $h_i$ is chosen with uniform probability from the interval $[-1, 1]$ and $u_i$ is chosen from $[0, 2]$, for each hysteron independently. Hysteron $j$ imposes a local field on hysteron $i$ equal to $\Jij s_j$ where the coupling strength $\Jij$ is taken to be symmetric ($\Jij = J_{ji}$) except where stated otherwise. The magnitude of each $\Jij$ (with $i \neq j$) is selected with uniform probability so that $|\Jij| \leq 1$. 

To capture the effect of the characteristic quadrupolar elastic deformations of rearranging soft spots, the signs of the $\Jij$ are dictated by the spatial configuration of the hysterons. Pairs that are $45^\circ$ off the shear direction have a frustrated coupling (antiferromagnetic, $\Jij, J_{ji} < 0$), whereas pairs along $0^\circ$ or $90^\circ$ have a cooperative coupling (ferromagnetic, $\Jij, J_{ji} > 0$). This rule assumes that all soft spots' displacement fields have approximately the same orientation and polarity relative to the direction of shear, which appears to be true broadly in experiments~\cite{Keim13,Keim14,Keim20,Chikkadi11}. 
% Even when we allow $\Jij \neq J_{ji}$, we always require $\Jij$ and $J_{ji}$ to have the same sign. 

Our simulations, available as an open-source Python package~\cite{code}, probe the system evolution under athermal, quasistatic, oscillatory driving between $-H_0$ and $+H_0$. We initialize the system with $H \ll -1$ and all hysterons negative ($s_i = -1$), and we evolve forward using an event-based method. % wherein we compute the field value for the next event (i.e., a hysteron changing states, or the applied field reaching an extremal value). 
Since flipping one hysteron may prompt a neighbor to flip, we wait for avalanches at fixed field until a stable state is reached; the hysteron farthest past its threshold is flipped first and all the local fields are updated between flips. %Avalanches proceed by flipping the hysteron farthest from its threshold and then updating all the local fields to check for further instability; 
In extremely rare cases where no stable state can be found or two flips are degenerate, the system is discarded. We continue driving until an absorbing state is reached where the dynamics repeat under further driving. %Figure~\ref{fig:2}a shows an example of a steady state with period $T=2$, where the four hysterons follow a repeating sequence of microscopic states that spans two periods of driving. 

To search for multiperiodic behavior efficiently given the couplings $\Jij$ and thresholds $H^\pm_i$, 
we note that increasing the driving amplitude $H_0$ will not change the dynamics until it is large enough to cause an additional hysteron to flip. Therefore a finite set of $H_0$ will exhaust all possible dynamics under symmetric driving.
To obtain this set,
for each of the $2^N$ possible states, we compute the two values of $H$ that bound the interval of stability for the state. 
We then sort the list of absolute values of these $H$, and take the midpoints between successive values as our set of $H_0$. 
% we obtain a set of amplitudes $\{H_0\}$ that exhausts all possible dynamics under symmetric driving.
%To search for multiperiodic behavior efficiently, we compute for each $\Jij$ all values of $H$ for which a hysteron \textit{could} flip. This is given by the set:
%\begin{equation}\label{eq:hflip}
%\{H\} = \bigl\{H^+_i -\sum_j \Jij s_j ,\; H^-_i -\sum_j \Jij s_j \bigr\},
%\{H\} = \bigl\{ | H^\pm_i - \sum_j \Jij s_j | \bigr\},
%\end{equation}
%considered over all $2^N$ possible states and all spin indices $i$. Driving the system at the midpoints between these values %(\textit{i.e.,} $\{(H_i+H_{i+1})/2\}$ and half of the smallest value) 
We perform a series of simulations starting with the smallest $H_0$ and continuing until any multiperiodic orbit is found. 
Such an ``amplitude sweep'' is likewise an efficient method to search for novel behavior in experiments.

\begin{figure*}[t]
\centering
\includegraphics[width=\textwidth]{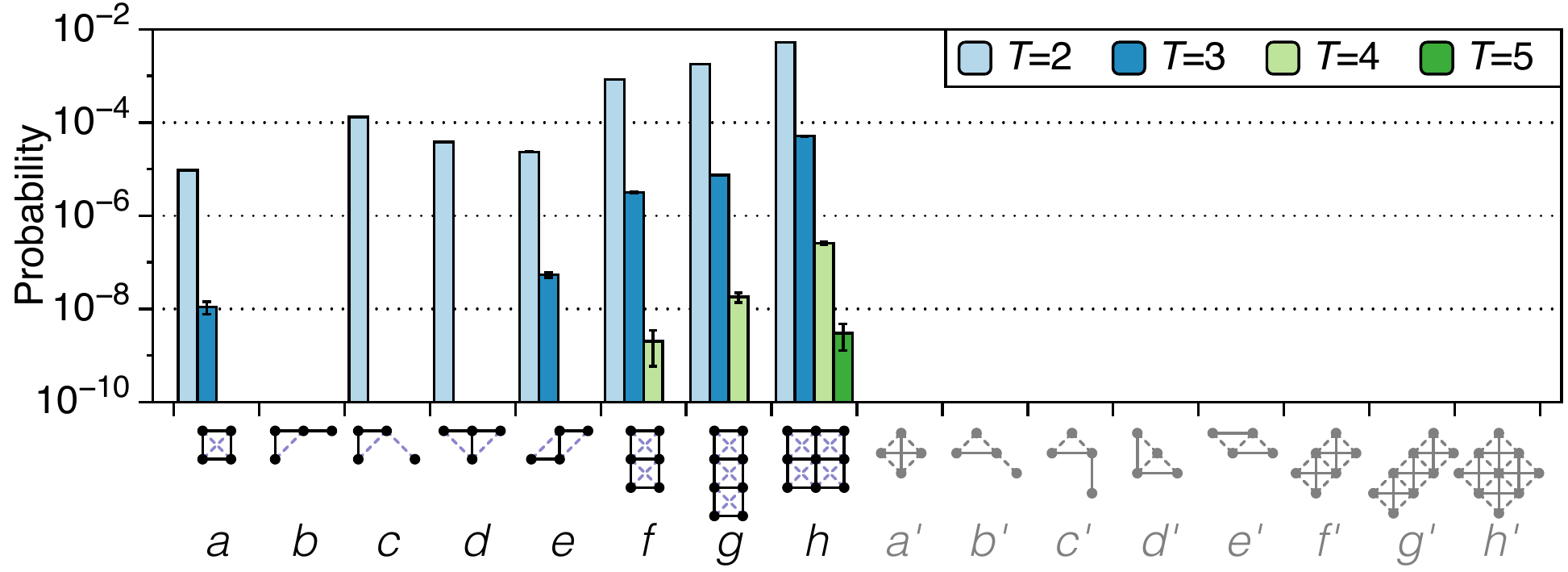}
\caption{
\textbf{How the prevalence of multiperiodicity depends on the spatial arrangements of the hysterons.} 
Probabilities of orbits with period $T=2, 3, 4, 5$, within the parameter space searched ($J_{ij} = J_{ji}$), for 8 arrangements of hysterons $a$--$h$. 
Arrangements $f$--$h$ exhibit a greater variety of periods and are consistent with an exponential decrease in probability as a function of the period (a straight line on these axes). In the diagrams along the $x$-axis, 
solid (dashed) lines represent cooperative (frustrated) interactions with $\Jij > 0$ ($\Jij  < 0$). For the complementary arrangements $a'$--$h'$, no multiperiodic behavior was found. Error bars represent 68\% confidence intervals; upper bounds on zero probabilities (not shown) are $1.3 \times 10^{-9}$~\cite{Feldman98}.
}
\label{fig:3}
\end{figure*}

\subsection*{Comparing arrangements of hysterons} 
Figure~\ref{fig:2} shows an example of a multiperiodic orbit that is achieved for $N=4$ hysterons arranged in a square. The system cycles through eight states over two driving periods, repeating this sequence indefinitely thereafter. This is just one possible $T=2$ orbit for this spatial arrangement of $N=4$ hysterons; it occurs with probability $P = 8.37 \times 10^{-6}$ (allowing permutation of hysterons and inversion of the $H_i^\pm$). 

Figure~\ref{fig:3} shows the prevalence of multiperiodicity in this and other compact arrangements of hysterons. The arrangements labelled $a$--$e$ show all the unique configurations where $N=4$ hysterons are placed within a $2\times 3$ lattice that is oriented with the shear direction (up to reflections and rotations by 90 degrees, which do not change the interactions). As before, interactions are between all nearest-neighbor pairs. Arrangement $c$ has the highest probability of $T=2$ among this set. 
These arrangements are some of the simplest ones eliciting multiperiodicity in our model. 

Arrangements $f$--$h$ in Fig.~\ref{fig:3} show the increasing prevalence of multiperiodicity for $N=6$, $8$ and $9$ hysterons on a square lattice. Arrangement $h$ has $P = 5.3 \times 10^{-3}$, so that if a macroscopic amorphous solid has 20 of these configurations, it will have a $\sim$10\% chance of multiperiodicity. Notably, in contrast to the observed behavior of amorphous systems of many particles~\cite{Regev13, Nagamanasa14, Lavrentovich17, Regev17}, small clusters of soft spots reach periodic orbits after very few cycles: for arrangement $h$, despite the space of $2^9$ states, the longest observed transient before a (multiperiodic) limit cycle was just 3 cycles, and it occurred in just 1 out of $10^7$ systems.

Remarkably, when the lattice is rotated by 45$^\circ$ (exchanging cooperative and frustrated interactions, i.e., $J_{ij} \rightarrow -J_{ij}$), no multiperiodic orbits are observed (arrangements $a'$--$h'$). 
% This holds even when we relax the symmetry $\Jij = J_{ji}$.
% Even when we allow $\Jij \ne J_{ji}$, these arrangements strongly suppress multiperiodic orbits by a factor of $\mathcal{O}(10^6)$ or more compared with $a$--$h$ --- for example, arrangement $g'$ yields just 4 multiperiodic orbits among $10^9$ cases.
% We did not investigate this curious observation further. 
This curious observation leads us to note another special property of $a'$--$h'$: if we assign a $+$ or $-$ state to any one hysteron, we can then work outward and assign states to all other hysterons, \emph{satisfying every interaction}. This is because these arrangements are portions of an antiferromagnetic lattice, with ordered ground states. While it is unclear why this property might suppress multiperiodic behavior, it could be a starting point for a deeper understanding of multiperiodicity generally.

Having demonstrated multiperiodic orbits in our simple model constructed from coupled hysterons, in the following sections we identify which attributes of the model are necessary for producing multiperiodicity.

\begin{figure}[t]
\centering
\includegraphics[width=0.5\textwidth]{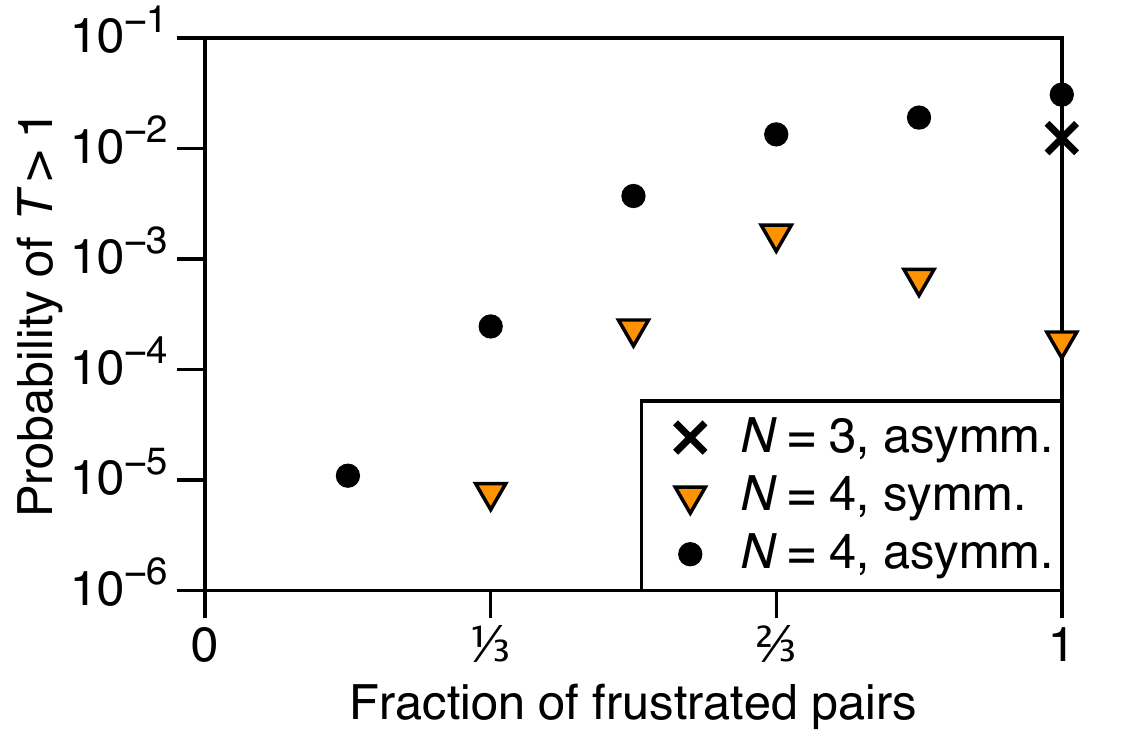}
\caption{
\textbf{Occurrence of multiperiodic behavior for few hysterons.} Probability is plotted as a function of the fraction of interaction pairs that are frustrated, for symmetric ($J_{ij} = J_{ji}$) and asymmetric interactions. 
%Notably, for $N=3$ with symmetric interactions, and for any configuration without frustration, no $T > 1$ was observed among $10^8$ systems. 
Error bars are smaller than the symbols.
Multiperiodic behavior becomes exponentially less common as the fraction of frustrated pairs is reduced from $2/3$ down to $0$, but with no observed multiperiodicity in all $10^8$ systems when there are no frustrated pairs. (A straight line on these axes corresponds to an exponential trend.) For N = 3 with symmetric interactions, we observed no multiperiodic orbits at all. 
}
\label{fig:small}
\end{figure}

\subsection*{Minimal number of hysterons} 
Empirically, we find that multiperiodic behavior is impossible for $N < 3$ hysterons. $N = 3$ hysterons with symmetric couplings also do not exhibit multiperiodic behavior. 
However, breaking the symmetry of at least one interaction pair ($J_{ij} \ne J_{ji}$) is enough to allow a $T=3$ orbit, if and only if all interactions are frustrated. 
Under these conditions, 
% \note{Allowing this asymmetry while still keeping the sign of $J_{ij}$ and $J_{ji}$ the same,}
we observe $T=3$ with $P= 4.67 \times 10^{-3}$, with a single unique sequence of states (see Supplemental Materials). 
We observe $T=2$ with $P = 7.80 \times 10^{-3}$, accounting for a variety of different sequences. 
%The resulting orbits with $T=3$ all follow a unique sequence (see SI Appendix) \note{[Is this sequence a distraction here? It is still in the SI too.]}
%\[++-, +--, +-+, --+, -++, -+-\] (allowing for permutation of the hysterons), with probability $(4.670 \pm 0.007) \times 10^{-3}$; a variety of $T=2$ orbits also arise. %with probability $(7.795 \pm 0.009) \times 10^{-3}$.

Asymmetric couplings in spin systems \emph{without} external cyclic driving \cite{Kinzel87,Hertz87,Bastolla97} have been studied before, but the physical meaning in a driven amorphous solid is unclear. One possible mechanism might be for soft spots to change states on different timescales, so that when the system is driven at finite frequency, a ``slow'' hysteron could fail to change in part of the cycle, even when in strict terms it is unstable.

\subsection*{Role of frustration} 
The observation that all interactions must be frustrated to elicit multiperiodic behavior for $N=3$ prompts us to further investigate the role of frustration. 
In Fig.~\ref{fig:small} we vary the fraction of interaction pairs that are randomly chosen to be frustrated ($J_{ij},J_{ji} < 0$), and we plot the prevalence of multiperiodicity under these conditions. 
There is a clear trend across all the data: Multiperiodic behavior becomes exponentially more scarce as the fraction of frustrated pairs is reduced from $2/3$ down to $0$. 
%For $N=4$, the probability of multiperiodicity decreases as the fraction of frustrated pairs is reduced below $2/3$, for both symmetric and asymmetric interactions. 
In all cases, the probability 
%of multiperiodicity 
is identically zero in the absence of frustration, a result we have checked up to $N=7$. 
%JDP: I reworked the last 3 sentences to make the central point more clear.
%
%Remarkably, Fig.~\ref{fig:small} indicates that for $N=4$, symmetric couplings with $2/3$ of pairs frustrated generally allow $T > 1$---confirming that arrangement $a'$ in Fig.~\ref{fig:3}, with no multiperiodic orbits, is a special case. 
%Remarkably, Fig.~\ref{fig:small} indicates that for $N=4$ with $2/3$ of pairs frustrated, $T > 1$ is allowed generally.
%---whereas simulations of arrangement $a'$ in Fig.~\ref{fig:3} find no multiperiodic orbits for symmetric or asymmetric interactions, confirming this arrangement is a special case. 
Figure~\ref{fig:small} also confirms that the topology of frustrated and cooperative interactions can be just as important as their number: arrangement $a'$ in Fig.~\ref{fig:3} has $N=4$ and $2/3$ of pairs frustrated, and yet we find no multiperiodic orbits for that specific topology, for either symmetric or asymmetric interactions.

%\note{This is a great place to point out that it's also a special case for asymmetric interactions.}

\subsection*{Multiperiodicity from non-hysteretic elements}
The above results show how coupled hysterons can produce multiperiodic orbits. 
We now show that hysteresis of the elements is in fact not a necessary ingredient for multiperiodicity. 
%It is easy to see that individual non-hysteretic elements may become hysteretic when coupled together---for example, two elements with $H^+ = H^- = 0$ and $J_{12} = J_{21} > 0$. More generally, 
In the absence of hysteresis and when $-1 < \Jij=J_{ji} < 1$, our model of an amorphous solid reduces to a spin glass where each soft spot corresponds to an Ising spin, governed by the Hamiltonian: \begin{equation}\label{eq:hamiltonian}
\mathcal{H} = -\frac{1}{2}\sum_{i\neq j} \Jij s_i s_j - H\sum_i s_i \ .
\end{equation}
We verified this by writing separate code for such a spin glass and comparing the results with our coupled hysteron code with zero hysteresis. Deutsch \& Narayanan \cite{Deutsch03} reported multiperiodic orbits in such spin glasses with as few as $5$ spins, although they focused on larger systems ($N \geq 64$). We now elucidate the conditions for multiperiodicity with $N=5$, under additional conditions that simplify the interactions even further: all the spin couplings are antiferromagnetic ($\Jij \leq 0$), and one or more of the couplings are randomly set to zero. 

With 4 couplings set to zero, no multiperiodic orbits were observed in $10^6$ systems. With 3 couplings set to zero, out of $10^7$ systems we observe multiperiodicity in $1,932$---all with period $T=3$ and a unique topology of interactions. 
This topology is shown in Fig.~\ref{fig:4}a, and in the inset to Fig.~\ref{fig:4}b as a portion of a triangular lattice. 
Without loss of generality, we break the mirror symmetry by requiring $|J_{34}|<|J_{01}|$ when spins are indexed left to right. This leads to an additional remarkable uniqueness: at the smallest $H_0$ for multiperiodicity in each system, there is a unique and highly symmetric steady-state orbit (see Supplemental Materials). %The sequence is: $[\da\ua\da\ua\da]$, $\da\ua\da\ua\ua$, $(\ua\ua\da\ua\ua)$, $\ua\da\da\ua\ua$, $[\ua\ua\da\ua\ua]$, $\ua\da\da\ua\ua$, $(\ua\da\da\ua\da)$, $\ua\da\ua\ua\da$, $[\ua\da\da\ua\da]$, $\ua\da\ua\ua\da$, $(\ua\da\ua\ua\ua)$, $[\ua\da\ua\da\ua]$, $\ua\da\ua\da\da$, $(\da\da\ua\da\da)$, $\da\ua\ua\da\da$, $[\da\da\ua\da\da]$, $\da\ua\ua\da\da$, $(\da\ua\ua\da\ua)$, $\da\ua\da\da\ua$, $[\da\ua\ua\da\ua]$, $\da\ua\da\da\ua$, $(\da\ua\da\da\da)$, 
%where unstable states are denoted with parentheses, and brackets denote the states at $\pm H_0$. 
%Although it may at first look cumbersome, there are several simple aspects: 
%(i) the sequence has an inversion symmetry wherein advancing $1.5$ cycles from any state inverts every spin,
%, which is evident by comparing the two lines showing $1.5$ periods each: each state is inverted with respect to the state above/below it.  
%(ii) the fourth spin is up for $1.5T$ and then down for $1.5T$, and (iii) the first spin behaves the same but delayed $2$ events. 

\subsection*{Regions in parameter space} The evolution of this spin-glass model is deterministic given the coupling strengths $\Jij$, an initial condition, and a driving protocol. Working in the reverse direction, a sequence of states may be mapped back to a region of the (high-dimensional) space of $\Jij$ that can give this sequence; here a subset of the unit hypercube $[-1,0]^n$, where $n=7$ is the number of nonzero couplings. 
Proceeding in this manner, we find a set of $10$ inequalities among the $\Jij$ that bound the region of parameter space corresponding to this $T=3$ orbit, which we list in the Supplemental Materials. The volume of this high-dimensional polygon (i.e., polytope) is found to be $1.86 \times 10^{-4}$. 

To convert to a probability for multiperiodicity, we multiply this volume by $2$ for the indexing degeneracy we lifted, and by $1/2$ to account for the probability of obtaining the correct network topology. 
The latter factor may be found by noting that at least two of the three removed edges must share a vertex and enumerating the remaining cases. 
Thus, the above value is precisely the predicted probability of multiperiodic behavior for some $H_0$. It agrees with how often we observe $T=3$ in our simulations: $P=(1.93 \pm 0.04) \times 10^{-4}$.

\begin{figure}[t]
\centering
\includegraphics[width=0.5\textwidth]{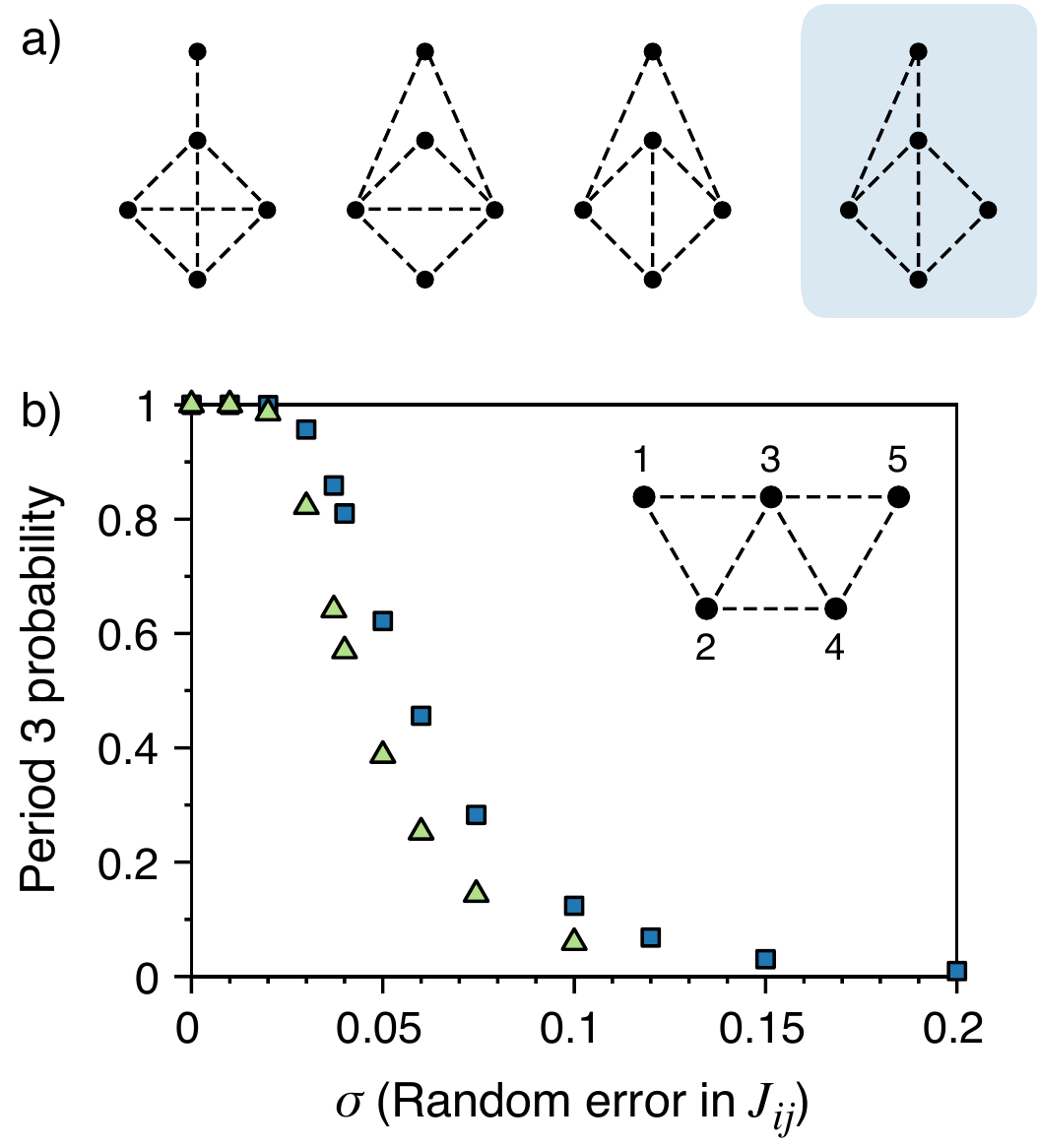}
\caption{
\textbf{Multiperiodicity from frustrated non-hysteretic elements.} 
\textbf{(a)} The four distinct graphs on $N=5$ vertices with 3 missing edges. 
Multiperiodic orbits were found only in the rightmost graph, redrawn in (b) as a portion of a triangular lattice. 
%Starting with a complete graph on $5$ vertices and removing $3$ edges at random, these topologies are obtained with the following probabilities (from left to right): $1/4$, $1/8$, $1/8$, $1/2$. 
\textbf{(b)} Probability of period 3 near the Chebyshev center of the period-3 polytope for the spin model in the inset, with $J_{12}<J_{45}$ to lift a degeneracy. Squares: Probability of falling within the polytope for $10^6$ Gaussian-distributed points around the Chebyshev center, while keeping $J_{14}=J_{25}=J_{15}=0$. Triangles: All $\Jij$ are given random errors. In this case the 7-dimensional description does not apply; one would need to characterize a distinct 10-dimensional polytope. Instead, $10^4$ simulations are run for each $\sigma$.
}
\label{fig:4}
\end{figure}

Such a detailed characterization of the high-dimensional phase space of the $\Jij$ is useful for designing systems with robust multiperiodic behavior. For instance, we can compute \cite{pypoman} a Chebyshev center for this polytope---a point that is farthest from its faces, which can thus withstand the largest possible errors in $\Jij$ while remaining multiperiodic. We find a Chebyshev center at %$(\Ja,\Jb,\Jc,\Jd,\Je,\Jf,\Jg) = -(0.926, 0.519, 0.703, 0.703, 0.370, 0.278, 0.297)$,
\begin{equation}
    J = \begin{bmatrix}
0 & -0.926 & -0.370 & 0 & 0 \\
-0.926 & 0 & -0.519 & -0.278 & 0 \\
-0.370 & -0.519 & 0 & -0.703 & -0.297 \\
0 & -0.278 & -0.703 & 0 & -0.703\\
0 & 0 & -0.297 & -0.703 & 0 \\
\end{bmatrix}
\label{eqn:spincenter}
\end{equation}
which is the center of a hypersphere of radius $0.074$ that lies entirely within the polytope.  
We report these coordinates to illustrate that our method can give precise quantitative information about finite regions of phase space that share a common orbit. 
At these coordinates, $T=3$ is attained for any $H_0$ in the range: $1 < H_0 < 1.685$ (see Supplemental Materials).
%Figure~\ref{fig:4} shows that this system can tolerate normally-distributed errors in the $\Jij$ of several percent while retaining $T=3$ behavior. 
For the general case of normally-distributed errors in the $\Jij$, 
Fig.~\ref{fig:4}b shows that the probability of $T=3$ remains high for a standard deviation $\sigma$ up to several hundredths. 
Thus, the low probability of multiperiodicity in this system stems from the enormity of the parameter space, rather than a need for fine-tuning.  

This same methodology---starting from an orbit and working backwards to a region of parameter space---also applies to our model of interacting hysteretic soft spots. For example, setting $H_0 = 1$, a Chebyshev center for the orbit in Fig.~\ref{fig:2} is 
\begin{subequations}
    \begin{align}
    J &= \begin{bmatrix}
0 & -0.552 & 0.081 & 0.081 \\
-0.552 & 0 & 0.670 & 0.280 \\
0.081 & 0.670 & 0 & -0.571 \\
0.081 & 0.280 & -0.571 & 0
\end{bmatrix} \\
H^+ &= \begin{bmatrix} 0 & -0.105 & 0.762 & 0.953 \end{bmatrix} \\
H^- &= \begin{bmatrix} -0.809 & -0.220 & -0.856 & -1.047 \end{bmatrix}
\end{align}
\label{eqn:center4}
\end{subequations}
which is a distance 0.081 from the nearest face.
For the unique $T=3$ orbit with $N=3$ hysterons, a Chebyshev center is 
\begin{subequations}
    \begin{align}
    J &= \begin{bmatrix}
0 & -0.586 & -0.172 \\
-0.172 & 0 & -0.586 \\
-0.586 & -0.172 & 0
\end{bmatrix} \\
H^+ &= \begin{bmatrix} 0.828 & 0.828 & 0.828 \end{bmatrix} \\
H^- &= \begin{bmatrix} -0.828 & -0.828 & -0.828 \end{bmatrix}
\end{align}
\end{subequations}
which is a distance 0.338 from the nearest face. 
% Note that this radius means that we can allow pair(s) of couplings to have opposite signs (i.e., relaxing $J_{ij} \le 0$), and the $T=3$ orbit is still possible! We have yet to investigate whether this works in the simulation. Also, this is not of strong interest because such interactions seem unphysical.
Note the high degree of symmetry at this Chebyshev center: Each hysteron has identical $H^+$ and $H^-$, with identical asymmetric couplings that set up a clear chirality in the system. 
In the Supplemental Materials, we further characterize all of the above polytopes and list the inequalities that bound them.

\section*{Discussion}
%To our knowledge, multiperiodicity is the first global, steady-state behavior of the system that requires frustrated interactions to explain. %---a behavior that should be unmistakable in experiments, for which we offer specific guidance and a testable hypothesis. 

We have shown how multiperiodicity can arise from the interactions of a small number of localized soft spots with simple, physically-motivated interactions. Previous studies of this behavior using molecular dynamics simulations did not consider localization to soft spots~\cite{Schreck13,Regev13,Royer15,Kawasaki16,Lavrentovich17,Nagasawa19,Yeh20}, 
while previous attempts to understand it using simplified models \cite{MunganTerzi19, MunganWitten19,Szulc20} did not pursue a microscopic picture of the system, e.g., of the sequence or spatial structure of rearrangements. 
In this work, by focusing on small systems, probing the effect of the spatial structure of the elements, and using an amplitude sweep for the driving field, we have provided a concrete and thorough foundation for addressing the origin of multiperiodicity in amorphous solids, where its robust appearance in simulations has not been well understood. 
While we have not specialized to particular %non-uniform 
distributions of parameter values that are a subject of current research~\cite{Nicolas18,Richard20}, our general model is nevertheless able to capture a more detailed aspect of the multiperiodicity found in molecular dynamics simulations: We observe an approximately exponential decay of probability with the period of the limit cycle, $T$ %(most prominently in arrangement $h$ in
(e.g., in arrangements $f$--$h$ in Fig.~\ref{fig:3}), a trend that was reported by Lavrentovich \textit{et al.}~\cite{Lavrentovich17} in simulations on jammed solids.
The findings of Lavrentovich \textit{et al.}~%\cite{Lavrentovich17} 
that multiperiodicity may be associated with an unjamming transition prompts the question of the role of soft-spot interactions in this critical transition. 

Our results show that frustrated interactions are always necessary for multiperiodic behavior. This comports with existing theory about the random-field Ising model~\cite{Sethna93}
which showed that without frustration, it supports return-point memory---a behavior that precludes a multiperiodic response. 
% While a generic set of \emph{sufficient} conditions for multiperiodicity in our model remains to be found, evidence that the behavior is impossible in any portion of an antiferromagnetic lattice (Fig.~\ref{fig:3} $a'$--$h'$) may be a starting point.
These findings suggest that multiperiodicity should be taken as a conspicuous 
signature of frustration---a counterpart to the yielding and shear-banding behaviors that are often attributed to cooperative interactions~\cite{Maloney04, Regev15, Nicolas18}.
%Cooperative interactions between soft spots are invoked to explain yielding and shear-banding phenomena in amorphous solids. 
%To our knowledge, multiperiodicity is the first global, steady-state behavior of the system that requires frustrated interactions to explain. %---a behavior that should be unmistakable in experiments, for which we offer specific guidance and a testable hypothesis. 

Our results also offer guidance to experiments searching for multiperiodicity in amorphous solids. Because interactions among soft spots are crucial, strain amplitudes should be large enough to ensure a high density of switching soft spots, but small enough to allow a periodic steady state---consistent with results of prior simulations that we can now rationalize with our model. Such experiments also promise to reveal the role of soft spot interactions near yielding~\cite{Mungan19,Nicolas18}, and to probe the limits of the return-point memory behavior that is incompatible with frustration~\cite{Sethna93,Paulsen19,Keim19,Mungan19,Keim20}. However, experiments must overcome measurement error and a high susceptibility to mechanical noise in this regime~\cite{Keim14,Keim20}. We have shown that relatively few soft spots are sufficient for multiperiodic behavior, so that localized clusters of soft spots may be the dominant way that multiperiodicity emerges in large systems. %, as interactions between soft spotso quadrupolar strain fields decay strongly with distance. 
Dividing observations of a large experimental system into regions of $\mathcal{O}(10)$ soft spots could thus enhance sensitivity to multiperiodic orbits, while rejecting the effects of mechanical noise or initial conditions playing out elsewhere. Furthermore, it would test the hypothesis that multiperiodic behavior is highly localized, rather than being a strictly emergent behavior spread out among many interacting particles. 
Combinations of small groups with incommensurate periods may be a way for longer-period orbits to arise. %\note{Royer hints [very obliquely] at this idea in their SI but does not come out and say it.} 

We have also shown that specific multiperiodic behaviors among spins and hysterons correspond to convex regions in high-dimensional parameter space, bounded by systems of inequalities. This both serves as an additional check of our modeling, and paves the way for the rational design of systems with these behaviors---for example, as the basis for a digital counter. 
Most promising are the $N=5$ spin configuration (Fig.~\ref{fig:4} inset) and the $N=3$ and $N=4$ hysteron configurations (Figs.~\ref{fig:2}--\ref{fig:small}), each of which is conducive to a real-space physical implementation, with network topology and bond strengths that might be realized in the lab.

\noindent \textbf{Supplementary Material} accompanies this paper at {\small {\tt http://www.scienceadvances.org/}}.

\section*{Materials and Methods}
\subsection*{Details for Fig.~\ref{fig:1}}
The experimental particle trajectories and micrograph used for Fig.~\ref{fig:1} were obtained using methods described in Ref.~\cite{Keim20}, by cyclically shearing a monolayer of bidisperse polystyrene particles adsorbed at an oil-water interface. Because these particles exhibit long-range electrostatic repulsion~\cite{Masschaele10}, the material is a disordered, frictionless soft solid. We shear each sample between parallel boundaries that are 1.5~mm apart and 18~mm long; the material extends far beyond the open ends of this working sample. We image an approx.~1.4~$\times$~1.9~mm region within the working sample. In the Supplementary Materials Fig.~S1, we show micrographs corresponding to Fig.~\ref{fig:1}(a, b).

Each material is prepared by combining small and large sulfate latex microspheres (Invitrogen) in suspension, in roughly equal number, and dispersing them at an oil-water interface~\cite{Keim20}. The small and large particles in Fig.~\ref{fig:1}a  have average diameters 3.8 and 5.2~$\mu$m, and the particles in Fig.~\ref{fig:1}(b, c) have average diameters 3.5 and 5.4~$\mu$m, though it is ultimately each particle's electric dipole strength that determines its effective size in the packing~\cite{Masschaele10}. Although aggregates of several particles can form during the preparation process, they do not seem to be strongly correlated with the locations of soft spots.

%JDP: we lead off with "To BLANK, ..." in the next paragraph, and also at other points in the paper.
%To obtain 
We obtained the plotted displacements (panels a and b)
%, we compared
by comparing the position of each particle at two different times, subtracting the average motion of the region of surrounding material with radius 16.5$a$, where $a$ is the mode of the interparticle distance, determined from the pair correlation function $g(r)$~\cite{Keim14,philatracksv02,Keim20}. We chose times when the shear strain $\gamma = 0$ (the midpoint of shearing), one full cycle apart. Panel a shows displacements in a portion of the system upon switching from strain amplitude 0.038 to 0.055. Panel b shows displacements in a different experiment, upon switching from strain amplitude 0.045 to 0.050.

To obtain the plotted trajectory loops (panel c), we used positions over a full cycle of shearing at strain amplitude 0.055. Rather than subtracting the average motion within the region shown, we subtract the motion of a set of particles centered $\sim$35~$\mu$m below this region, so that the particles in the field of view appear to be displaced horizontally by the global shearing motion.

%\subsection*{Probability to obtain each topology in Fig.~\ref{fig:4}a} --- In the main text, we consider $N=5$ spins with three coupling strengths $\Jij$ randomly set to zero. This process always yields one of the four network topologies in Fig.~\ref{fig:4}a, but with unequal probabilities. To determine them, note that at least two of the three removed edges must share a vertex. It is then straightforward to enumerate the $8$ possible locations for the remaining removed edge. The probabilities are thus: $1/4$, $1/8$, $1/8$, $1/2$ (left to right in Fig.~\ref{fig:4}a).

\subsection*{Inequalities for regions of parameter space} 
For a system of spins, inequalities that bound regions of parameter space may contain only the parameters $H_0$ and $J_{ij}$ as variables.
To generate such inequalities from a sequence of states, we follow a method that parallels our simulation algorithm. Two examples illustrate our approach. We first consider a spin $i$ that flips to the $+$ state as $H$ is increased. At this instant the spin has become marginally unstable, so that
\begin{equation}
    H + \sum_{j \ne i} J_{ij} s_j = 0.
    \label{eq:flipfield}
\end{equation}
At this same instant the other spins are stable, since otherwise they would have flipped before spin $i$ did. For instance, if spin $k \ne i$ is in the $-$ state,
\begin{equation}
    H + \sum_{j \ne k} J_{kj} s_j < 0
    \label{eq:stablefield}
\end{equation}
where we use the previous states $s_j$ of the spins before spin $i$ flipped. Substituting Eq.~\ref{eq:flipfield} into Inequality~\ref{eq:stablefield} yields an inequality that contains only the unknowns $J_{ij}$, as desired.
As a second example, we imagine that the flipping of spin $i$ causes another spin $l$ to flip immediately (an avalanche). This tells us not only that spin $l$ is unstable at the same value of $H$ given by Eq.~\ref{eq:flipfield}, but that at that instant, it is farther past its threshold of stability than every other spin.
The avalanche ends when all spins are stable; this observation leads to further inequalities by again combining Eqs.~\ref{eq:flipfield} and \ref{eq:stablefield} (where Inequality~\ref{eq:stablefield} is flipped for spins in the $+$ state).
A similar method applies to a system of hysterons, with $H^+_i$ and $H^-_i$ as additional unknowns on the righthand side of Eqs.~\ref{eq:flipfield},\ref{eq:stablefield} as needed. In general, additional inequalities are needed to denote that $|H| \le H_0$ at all times.

The resulting inequalities define a high-dimensional polygon (polytope). We use the Python package \textit{pycddlib} (based on \textsc{CddLib}) to remove any redundant inequalities and the \textit{pypoman} package~\cite{pypoman} to compute Chebyshev centers. The full sets of inequalities for select orbits, further characterizations of the polytopes, and checks of the inequalities against our simulation results are given in the Supplemental Materials.

\subsection*{Period-3 polytope volume in the spin model} 
To compute the volume of the period-3 polytope for $N=5$ spins, we first convert from a set of inequalities to a set of vertices, using the Python package \textit{pycddlib}. %, a \textsc{Python} wrapper for Komei Fukuda's \textsc{CddLib}. 
We then compute the volume of the convex hull of these points with the SciPy module \textit{spatial.ConvexHull}. We find it to be $1.863 \times 10^{-4}$. Measuring this volume using Monte Carlo integration with $10^8$ points gives consistent results: $ (1.857 \pm 0.010) \times 10^{-4}$. The full set of inequalities defining the polytope, and the 14 vertices they define, are given in the Supplemental Materials. 

\subsection*{Organizing the hysteron simulation orbits} 
Comparing orbits lets us meaningfully group and count systems with equivalent orbits. We represent each simulation's output as a directed cyclic graph of states, and manipulate it with the NetworkX package~\cite{Hagberg08}. We obtain the orbit by extracting the longest simple cycle in this graph. This removes trivial excursions: for instance, a system may transition from state $+-+$ to $+++$ as $H$ as increased, and then return directly to $+-+$ as $H$ is decreased; we generally find many other randomly-generated systems in which this excursion is missing. To compare these extracted orbits, we then account for all possible permutations of hysterons' identities, reversal of the sequence, and inversion of the system (exchanging all the $+$ and $-$ states).
%, which is equivalent to transforming the system by $H^+_i \to -H^-_i$ and $H^-_i \to -H^+_i$.

% Your references go at the end of the main text, and before the
% figures.  For this document we've used BibTeX, the .bib file
% scibib.bib, and the .bst file Science.bst.  The package scicite.sty
% was included to format the reference numbers according to *Science*
% style.

\bibliography{references}
\bibliographystyle{ScienceAdvances}

\noindent \textbf{Acknowledgements:} 
% Acknowledgments should be gathered into a paragraph after the final numbered reference. This section should also include 
% * complete funding information, 
% * a description of each author's contribution to the paper, 
% * a listing of any competing interests of any of the authors (all authors must also fill out the Conflict of Interest form), and, 
% * a section on data and materials availability, information about the location of the data if not included in the paper, including **accession numbers** to any data relating to the paper and deposited in a public database.
%
We thank Chloe W. Lindeman and Sidney R. Nagel for bringing to our attention the connection of the exponential decay in Fig.~\ref{fig:3}f-h with previous simulations on jammed solids. \\
\noindent \textbf{Funding:} This work was supported by National Science Foundation Grant No. DMR-1708870 (N.C.K.). \\
\noindent \textbf{Author Contributions} N.C.K.\ and J.D.P.\ designed and performed research and wrote the paper.\\
\noindent \textbf{Competing Interests} The authors declare that they have no competing financial interests.\\
\noindent \textbf{Data and materials availability:} Essential source code is published online at \\ {\small \tt https://github.com/nkeim/hysteron}. All data needed to evaluate the conclusions in the paper are present in the paper and/or the Supplementary Materials.

% For your review copy (i.e., the file you initially send in for
% evaluation), you can use the {figure} environment and the
% \includegraphics command to stream your figures into the text, placing
% all figures at the end.  For the final, revised manuscript for
% acceptance and production, however, PostScript or other graphics
% should not be streamed into your compliled file.  Instead, set
% captions as simple paragraphs (with a \noindent tag), setting them
% off from the rest of the text with a \clearpage as shown  below, and
% submit figures as separate files according to the Art Department's
% instructions.

%\clearpage

%\noindent {\bf Fig. 1.} Please do not use figure environments to set
%up your figures in the final (post-peer-review) draft, do not include graphics in your
%source code, and do not cite figures in the text using \LaTeX\
%\verb+\ref+ commands.  Instead, simply refer to the figure numbers in
%the text per {\it Science\/} style, and include the list of captions at
%the end of the document, coded as ordinary paragraphs as shown in the
%\texttt{sciadvfile.tex} template file.  Your actual figure files should
%be submitted separately.
%

\end{document}